\documentclass[conference]{IEEEtran}
\IEEEoverridecommandlockouts
\usepackage{cite}
\usepackage{amsmath,amssymb,amsfonts}
\usepackage{algorithmic}
\usepackage{graphicx}
\usepackage{textcomp}
\usepackage{xcolor}
\usepackage{multirow}
\usepackage{array}
\def\BibTeX{{\rm B\kern-.05em{\sc i\kern-.025em b}\kern-.08em
    T\kern-.1667em\lower.7ex\hbox{E}\kern-.125emX}}

\title{Utilizing Large Language Models to Identify Reddit Users Considering Vaping Cessation for Digital Interventions\\
}

\author{
\IEEEauthorblockN{
    Sai Krishna Revanth Vuruma\IEEEauthorrefmark{1},
    Dezhi Wu\IEEEauthorrefmark{1}\textsuperscript{,}\IEEEauthorrefmark{4},
    Saborny Sen Gupta\IEEEauthorrefmark{1},
    Lucas Aust\IEEEauthorrefmark{1},
    Valerie Lookingbill\IEEEauthorrefmark{1},  \\
    Caleb Henry\IEEEauthorrefmark{1},
    Yang Ren\IEEEauthorrefmark{1},
    Erin Kasson\IEEEauthorrefmark{2},
    Li-Shiun Chen\IEEEauthorrefmark{2},
    Patricia Cavazos-Rehg\IEEEauthorrefmark{2},
    Dian Hu\IEEEauthorrefmark{3},
    Ming Huang\IEEEauthorrefmark{3}\textsuperscript{,}\IEEEauthorrefmark{4}}   \\
    
\IEEEauthorblockA{
\IEEEauthorrefmark{1}
\textit{HI3 Tech Lab} \\
\textit{Department of Integrated Information Technology} \\
\textit{University of South Carolina}\\
Columbia, SC, USA \\}   \\

\IEEEauthorblockA{
\IEEEauthorrefmark{2}
\textit{School of Medicine} \\
\textit{Washington University}\\
St. Louis, MO, USA \\}  \\

\IEEEauthorblockA{
\IEEEauthorrefmark{3}
\textit{Center of Translational AI Excellence and Applications in Medicine} \\
\textit{McWilliams School of Biomedical Informatics} \\
\textit{University of Texas Health Science Center}\\
Houston, TX, USA \\}    \\

\IEEEauthorblockA{
\IEEEauthorrefmark{4}Corresponding Authors:
Dezhi Wu, dezhiwu@cec.sc.edu;
Ming Huang, Ming.Huang@uth.tmc.edu;
}
}

\usepackage{bibentry}

\usepackage{comment}
\usepackage{booktabs}

\usepackage{color}

\begin{document}
\maketitle

\begin{abstract}
The widespread adoption of social media platforms globally not only enhances users' connectivity and communication but also emerges as a vital channel for the dissemination of health-related information, thereby establishing social media data as an invaluable organic data resource for public health research. The surge in popularity of vaping or e-cigarette use in the United States and other countries has caused an outbreak of e-cigarette and vaping use-associated lung injury (EVALI), leading to hospitalizations and fatalities in 2019, highlighting the urgency to comprehend vaping behaviors and develop effective strategies for cession. In this study, we extracted a sample dataset from one vaping sub-community on Reddit to analyze users' quit vaping intentions. Leveraging large language models including both the latest GPT-4 and traditional BERT-based language models for sentence-level quit-vaping intention prediction tasks, this study compares the outcomes of these models against human annotations. Notably, when compared to human evaluators, GPT-4 model demonstrates superior consistency in adhering to annotation guidelines and processes, showcasing advanced capabilities to detect nuanced user quit-vaping intentions that human evaluators might overlook. These preliminary findings emphasize the potential of GPT-4 in enhancing the accuracy and reliability of social media data analysis, especially in identifying subtle users' intentions that may elude human detection.   
\end{abstract}


\section{Introduction}
Due to the ubiquity of social media platforms, over 4.7 billion users worldwide use them for connectivity, communications, news, and entertainment with a significant portion of the discourse related to health. The utilization of social media data emerges as a nascent source of public health information, offering novel insights into public health trends and enhancing the capabilities of public health surveillance. 

In recent years, the United States has witnessed a significant surge in the popularity of vaping or e-cigarette use, leading to a notable rise in cases of e-cigarette and vaping use-associated lung injury (EVALI) that caused hospitalizations and fatalities in 2019. Previous research studies have leveraged popular Twitter and Reddit social media data for public surveillance of health topics. In particular, two previous social media vaping studies \cite{wu2022vapetwitter, KASSON2021104574} used topic modeling and sentiment analysis along with clinical insights to show social media users might benefit from digital intervention programs for vaping cessation and potentially for clinicians to employ proactive outreach strategies to engage vaping patients for patient education and treatment on social media platforms. Moreover, in some other recent studies on e-cigarette or vape use, many users reported symptoms of Vape Dependence\cite{soule2020cannot}, where they cannot go for long periods of time with out craving for a vape, with users commonly resorting to stealth-vaping\cite{yingst2019cigarette} in places that prohibit vaping. Further, researchers found that vape frequency was directly associated with perceived satisfaction while being indirectly associated with perceived danger\cite{kozlowski2017daily}. Alarmingly, a teenager vaping study \cite{stalgaitis2020vaping} has highlighted the youth as high risk population for targeted health communications. To tackle this pressing public health challenges, it is imperative to conduct further research into the analysis of vaping discourse on social media platforms, aiming to develop artificial intelligent (AI)-based approach to more efficiently and accurately identify these patients' social media vaping discourse behaviors and developing targeted vaping prevention and intervention programs for the youth population. 

In this study,  we aim to  employ and evaluate OpenAI's GPT-4 model and traditional  BERT-based language models and by comparing human evaluators on a data annotation task to identify vaping cession interests among Reddit users.  Our preliminary findings indicate GPT-4 model demonstrates better consistency in adhering to annotation guidelines and processes compared to human evaluators. Following the introduction, in the next section, we present the related work in this area, and then report our research methodology and preliminary study findings. Lastly, we draw study conclusion and propose future research directions.  
\section{Related Work}
In this technologically prudent era, social media platforms, such as Facebook, Instagram, Reddit, and Twitter, have emerged as pivotal spaces for self-expression and social interaction \cite{maceda2023classifying}. These platforms offer a massive data resource that reflects the perceptions of its users, and the extraction and analysis of these data extends an opportunity for different research studies \cite{maceda2023classifying, struik2021cigarette}. However, interpretation of natural language data requires deep contextual knowledge and understanding \cite{törnberg2023chatgpt4}.

Manual annotations of social media texts can be challenging for humans as the texts are short and informal and contain different socio-cultural opinions and perceptions \cite{maceda2023classifying, liyanage2023gpt}. Studies have shown that lack of contextual understanding can lead to incorrect annotations \cite{liyanage2023gpt}. Recent work has shown that data annotation process can be augmented with state-of-the-art Machine Learning (ML) algorithms and Natural Language Processing (NLP) models to great success \cite{stenetorp2012brat}. However, care must be taken while using these techniques in tasks requiring complex inferences as shown in \cite{törnberg2023chatgpt4, cheng2023gpt}. In this regard, advanced and intuitive Large Language Models (LLMs) such as OpenAI's Generative Pre-trained Transformer models GPT-3\cite{brown2020language} and GPT-4\cite{openai2024gpt4} and Meta's LLAMA 2 \cite{touvron2023llama} have gained popularity in recent years due to their proficiency in in-context learning where they outperformed traditional methods \cite{alhamed2024using, deng2023llms}. These models can generate quick results and are not susceptible to some of the limitations seen in human annotation \cite{pangakis2023automated}. 

Using the GPT models, the findings from \cite{ding2022gpt} show that while GPT-3 still has room for improvement when compared to human annotators, it is capable of annotating data from different domains. Another study used GPT-4 to label the political tweets collected from USA politicians \cite{törnberg2023chatgpt4} with better accuracy, reliability, and equal or lower bias of GPT-4 than the expert classifiers.
\section{Methodology}
Reddit is a free, popular social media platform where users can take part in discussions and share content. It facilitates conversations between like-minded users through dedicated communities called subreddits. Each subreddit is focused on a specific topic or interest shared amongst it's subscribers.

r/QuitVaping is a subreddit where users come together to motivate each other to quit vaping. With around 40,000 subscribers, it is the largest subreddit dedicated to help users quit vaping and other tobacco products.

\begin{figure*}[tb]
    \centering 
    \includegraphics[width=\textwidth]{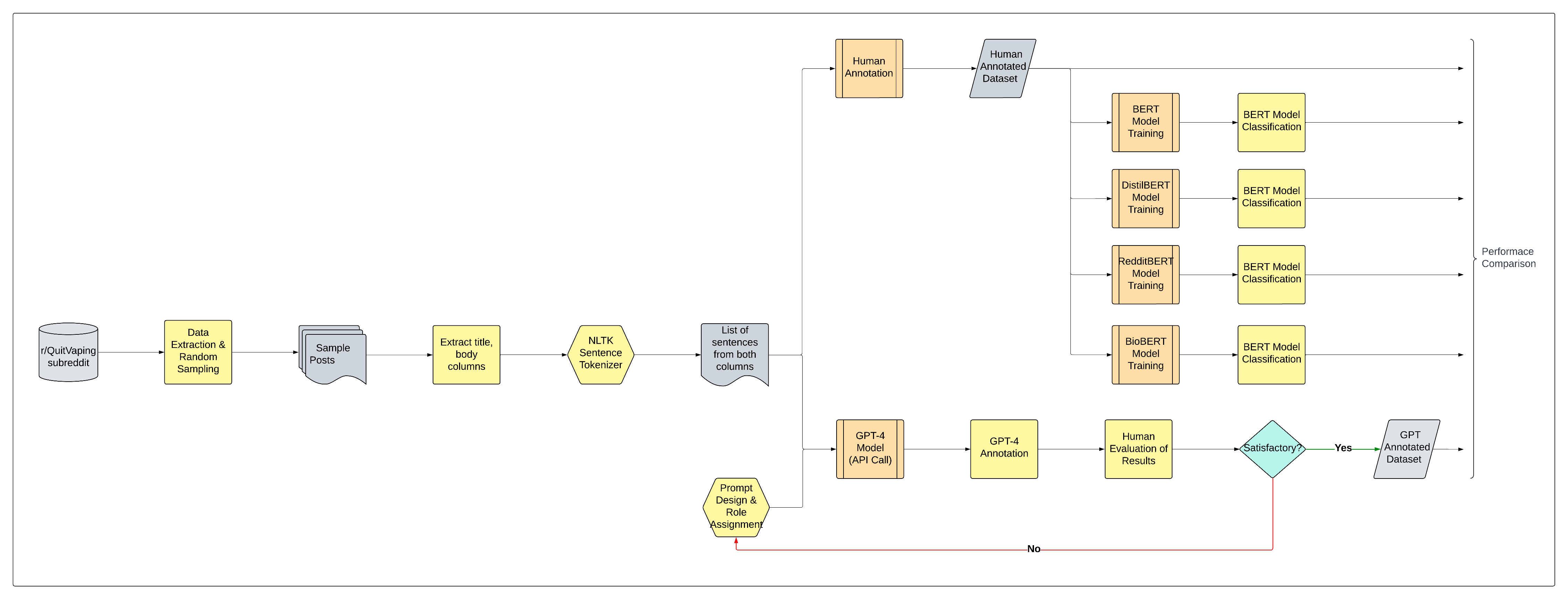}
    \caption{Workflow to evaluate GPT-4 and BERT-based language models}
    \label{fig_method}
\end{figure*}

The workflow adopted for this study is illustrated in Figure \ref{fig_method}. Each stage of the pipeline will be discussed in detail in the respective subsections. First the data is extracted and cleaned from Reddit, then sent to the human evaluators and the GPT-4 model for annotations. The human annotated dataset is then passed to pre-trained BERT models for training on a classification task. The performance of all three approaches is compared at the end to draw conclusions.

\subsection{Data Collection \& Preparation}
Reddit's Async PRAW API provides direct access to posts by subreddit through asynchronous execution. From the aforementioned r/QuitVaping subreddit, a total of 1000 posts made by users were extracted using the API. These posts ranged from users talking about their progress towards quitting vaping to users looking for help or motivation to quit or reduce vape use.

Out of these 1000 posts, approximately 120 of them were randomly selected to form a sample dataset. From each post in the sample dataset, two columns including title and body were extracted and broken down into sentences using the Sentence Tokenizer from the NLTK library \cite{nltk_sent_tokenize}. Table \ref{tab_post_bdwn} gives a statistical overview of the text in the title and body fields of each post in the sample dataset. \\
\begin{table}[tb]
\caption{ANALYSIS OF SENTENCES AND WORDS PER REDDIT POST}
\begin{center}
\begin{tabular}{|c|m{10em}|m{10em}|}
\hline
\textbf{Field Name} & \textbf{Average number of sentences per post} & \textbf{Average number of words per post} \\

\hline
\textbf{Title} & {1.07} & {6.56}\\

\hline
\textbf{Body} & {9.02} & {157.74}\\

\hline
\end{tabular}
\label{tab_post_bdwn}
\end{center}
\end{table}

\subsection{Human Annotation}
The vaping-related sentences from the extracted sample were assigned to two human evaluators for manual annotations related quit-vaping intention. The assigned labels from this step will serve as the ground truth while making performance comparisons among the ML approaches.

The annotators were tasked to label each sentence as 'Yes' or 'No' by following these guidelines to differentiate whether it is a quit-vaping related sentence:

\begin{itemize}
    \item \textbf{Yes:}
        \begin{itemize}
            \item User explicitly mentions their desire, plan, or decision to specifically quit vaping, e.g., \textit{ "I have been wanting to quit the vape for a while"}
        \end{itemize}
    \item \textbf{No:} 
    \begin{itemize}
        \item Sentences that do not mention vaping or quitting, e.g., \textit{"Sorry if it sounds overdramatic."}
        \item User explicitly states that they have already quit vaping, e.g., \textit{"5 days strong of cold turkey quitting all nicotine."}
        \item User describes a negative outcome of vaping, such as a negative health outcome, e.g., \textit{"It makes my mouth taste and smell weird."}
        \item User expresses negative feelings of or experiences with vaping, such as feelings of shame or embarrassment about vaping, e.g., \textit{"I work in healthcare and live a healthy lifestyle and am ashamed of vaping."} 
    \end{itemize}
\end{itemize}

The class distribution after human annotation was concluded can be found in Table \ref{tab_human_ann}.

\begin{table}[tb]
\caption{Class Distribution after Human Annotation}
\begin{center}
\begin{tabular}{|m{6em}|m{4em}|m{4em}|m{3em}|c|}
\hline
\textbf{Quit Vaping Intention} & \textbf{Training Data} & \textbf{Validation Data} & \textbf{Testing Data} & \textbf{Total}\\

\hline
\textbf{Yes} & {64} & {9} & {13} & {86} \\

\hline
\textbf{No} & {860} & {114} & {173} & {1147} \\

\hline
\textbf{Total} & {924} & {123} & {186} & \textbf{1233} \\

\hline
\end{tabular}
\label{tab_human_ann}
\end{center}
\end{table}

\subsection{Classification using pre-trained BERT models}
BERT\cite{devlin2019bert} has revolutionized language representations by introducing the concept of bidirectional attention in transformers. It served as the precursor to modern day Large Language Models (LLMs). Due to its architecture it can easily be fine-tuned to perform a multitude of tasks from question answering to natural language understanding. It can also be adopted to more specific use cases in social media studies such as COVID19 content analysis\cite{muller2023covid}, Fake News Detection\cite{kaliyar2021fakebert} or Hate Speech Detection\cite{mozafari2020hate} making it an ideal candidate for our purpose.

In order to fine-tune the base BERT model on our data, the sample dataset after human annotation was passed to the model as the input. The model was trained to classify the sentences from the title and body columns of the posts into 'Yes' or 'No' instances using the labels assigned by human evaluators as the target. The labelled dataset was broken down into train (75\%), validation (10\%) and test (15\%) sets using stratified sampling. Given the imbalance between the classes in the training data, we had to adjust the model weights to make if a fair classification. As shown in Table \ref{tab_human_ann}, the samples with label 'No' outnumber those with 'Yes' by about 13.5 times. To account for this lopsidedness, we assigned class weights for both classes and updated the loss function during training and evaluation to arrive at a more representative loss value.

Along with the base model, other pre-trained BERT models such as BioBERT\cite{Lee_2019}, RedditBERT and DistilBERT\cite{sanh2020distilbert} were also trained on the data using the same approach.

\subsection{GPT-4 Annotation}
In addition to using traditional transformer models, OpenAI's GPT-4 model was used to annotate the data. It is the latest model in the line of the popular Generative Pre-trained Transformers from OpenAI. It has shown remarkable capabilities in a multitude of domains, even clearing the bar exam according to a recent study\cite{katz2024gpt}. Recent studies indicate that GPT-4 can act as an alternative to human annotation in many diverse areas\cite{törnberg2023chatgpt4, cheng2023gpt, azad2023effectiveness}.

To test GPT-4's ability to annotate Reddit data from our sample dataset, we used OpenAI's API feature that lets users interact with the model of their choice via prompts. The model was tasked to do the following:
\begin{itemize}
    \item Follow the same guidelines that the human evaluators did for data annotation.
    \item Assign the label 'Yes' to a sentence if it indicates the speaker's intention to quit vaping right now or the label 'No' if it does not.
    \item Identify specific phrases from each sentence to justify your response.
    \item Provide an explanation of why that phrase does or does not indicate an intention to quit vaping.
\end{itemize}
\section{Results}
\subsection{Pre-trained BERT language models}
Four pre-trained variants of the BERT model were fine-tuned on a binary classification task using the labels from human annotation as the ground truth. The summary of the classification results on the test dataset for each model is shown in Table \ref{tab_bert_results}. Although the overall accuracy and recall suggest that the models are able to generalize predictions on unseen data, all four variants had issues identifying the positive class ('Yes' instances) on the test dataset. This can in part be owed to the fewer number of 'Yes' cases in the training data, even after the loss function was adjusted to better reflect the train dataset.

Out of the four BERT variants that were tested, BioBERT had the best classification metrics on the test dataset with an F1 score of 95. On the 'Yes' instances in the test data, it had the best recall of about 60\%. In comparison, the variant with the second best performance had a recall of 43\% on the positive instances.

DistilBERT on the other hand, had the lowest recall (0\%) for the positive class on the test dataset. It was unable to correctly classify any instance of the positive class. The base BERT model and RedditBERT were able to make inferences closer to the BioBERT model. However, their performance on the 'Yes' instances was not up to par with BioBERT.

\begin{table}[tb]
\caption{BERT Models' Performance on Classification}
\begin{center}
\begin{tabular}{|c|c|c|c|}
\hline
\textbf{Model} & \textbf{Accuracy} & \textbf{Recall} & \textbf{F1 Score} \\

\hline
\textbf{bert-base-uncased} & {94\%} & {95\%} & {94} \\

\hline
\textbf{distilbert-base-uncased} & {85\%} & {92\%} & {89} \\

\hline
\textbf{reddit-bert-text\_10} & {93\%} & {94\%} & {93} \\

\hline
\textbf{biobert-base-cased-v1.2} & {95\%} & {96\%} & {95} \\

\hline
\end{tabular}
\label{tab_bert_results}
\end{center}
\end{table}

\subsection{GPT-4}
While interacting with the API, two fields including role and prompt inform the model what its intended purpose is. 'Role' refers to the role of the model in this conversation while the 'prompt' field is used to provide input text to the model. We observed differences in model performance upon altering either of these two fields. For evaluating the performance of the GPT-4 model, we used the human annotated dataset as the ground truth for calculating metrics.

The model tends to perform better when it's role is clearly defined. We evaluated the model's performance with three different role definitions as shown in Table \ref{tab_gpt_roles} with the base definition being "You are an intelligent assistant". For the other two roles, we gave the model more context and some example labeled sentences in order to give it a better understanding of it's task. The model's recall went up from 48.1\% to 74.1\% when the role was more specific and had some examples for context.

However, the model still made a lot of false positive predictions with it's precision as low as 12.4\%. For example, the sentence \textit{"I have a big goal of affording an Apple Vision Pro at this Christmas and the money I spend on vaping would make that goal legit go up in smoke I want to look back on this post and know; nicotine was never your friend."} was marked as positive ('Yes') by GPT-4. Although the speaker talks about distancing themselves from vaping, there is no explicit or implicit mention of quitting vaping right now.

Subtle changes in words and sentence phrasing affect the way the model responds to a prompt. One instance of hallucination was noted when the model started evaluating every sentence in the context that it was related to vaping. For example, it marked the sentence "Quit!" as Yes and reasoned that the word quit indicated that the speaker wanted to quit vaping. Not only does that go against the annotation guidelines, but it shows yet another instance of Generative AI models inferring data that isn't there.

\begin{table}[tb]
\caption{GPT-4 Model Role Definitions}
\begin{center}
\begin{tabular}{|p{18em}|c|}
\hline
\textbf{Role} & \textbf{Change} \\

\hline
{You are an intelligent assistant} & {-} \\

\hline
{You are an intelligent assistant + Annotation guidelines used by human evaluators} & {added context} \\

\hline
{You are an intelligent assistant + Annotation guidelines used by human evaluators + Sample sentences with labels} & {added context, examples} \\

\hline
\end{tabular}
\label{tab_gpt_roles}
\end{center}
\end{table}

\subsection{Comparison}
Before drawing any conclusions, it is important to note the limitations of this study:
\begin{itemize}
    \item \textbf{Training Data:}
    \begin{itemize}
        \item \textbf{Size:} For this preliminary study, we used a relatively smaller dataset to train the BERT models. It is no secret that the models could have benefited with a larger training corpus.

        \item \textbf{Diversity:} All the posts were taken from the same subreddit community which may not be representative of the discourse on other subreddits.
    \end{itemize}
\end{itemize}

With a detailed and well-constructed prompt and precise role, GPT-4 demonstrated impressive capability to assist in social media data annotation. However, caution must be taken to evaluate the model's responses at each stage of the pipeline to ensure that it is not biased. Key findings from our comparison analysis indicate that the GPT-4  model was more consistent in following annotation guidelines than the human evaluators. It was able to detect implied quit vaping intentions that the human evaluators sometimes missed.
\section{Conclusion and Future Work}
With this preliminary study, we were able to compare the performance of OpenAI's GPT-4 model against traditional pre-trained BERT-based language models and human evaluators on a data annotation task to identify vaping cessation interests among Reddit users. We found that GPT-4 showed promising results, sometimes even outperforming the human evaluators. However, caution must be taken at each stage of the pipeline to minimize biases. 

In the future, we plan to further develop this study as follows:
\begin{itemize}
    \item \textbf{Larger Dataset:} Using the phrases and sentences that mention quit-vaping intentions, a much larger dataset can be collected through techniques such as semantic search.
    \item \textbf{Multi-label Classification:} This preliminary study only examined users' quit-vaping intentions at the sentence level due to the nature of work-in-progress. Given the complexity of the social media data, we plan to include more vaping-related labels such as negative health outcomes,  reducing vape use, etc.
    \item \textbf{Contextual Learning:} In this study, each sentence was evaluated individually without the added context of the post that it was taken from. Machine Learning models tend to perform better with more context, so we will expand this work  to contextual learning for the next-phase development.
\end{itemize}
\section*{Acknowledgements}
This research was generously funded by a NIH R34 grant (Grant No: CL040 155600 F1000 202 USCSP 10012339 1) and another research grant by the University of South Carolina (USC) (PI: Dr. Dezhi Wu, Grant No: 80002838).

\bibliographystyle{IEEEtran}
\bibliography{all4health}

\end{document}